\documentclass[preprint]{elsart}
\usepackage{graphicx}
\usepackage{amssymb}
\usepackage{multirow}

\begin{document}

\begin{frontmatter}

\title{Study of some two-body non-mesonic decays of $^4_\Lambda$He and
$^5_\Lambda$He}

\centering{FINUDA Collaboration}

\author[a,b]{M.~Agnello}, 
\author[c]{L.~Benussi}, \author[c]{M.~Bertani}, \author[d]{H.C.~Bhang},
\author[e,f]{G.~Bonomi}, \author[h,b]{E.~Botta},
\author[ii]{M.~Bregant},
\author[h,b]{T.~Bressani},
\author[b]{S.~Bufalino},
\author[k,b]{L.~Busso}, \author[b]{D.~Calvo}, \author[i,j]{P.~Camerini},
\author[ll]{B.~Dalena}, 
\author[h,b]{F.~De Mori},
\author[l,m]{G.~D'Erasmo},
\author[c]{F.L.~Fabbri}, \author[b]{A.~Feliciello}, 
\author[b]{A.~Filippi\thanksref{corresponding}},
\author[l,m]{E.M.~Fiore}, \author[f]{A.~Fontana}, \author[n]{H.~Fujioka},
\author[f]{P.~Genova}, \author[c]{P.~Gianotti}, \author[j]{N.~Grion},
\author[c]{V.~Lucherini}, \author[h,b]{S.~Marcello}, 
\author[o]{N.~Mirfakhrai}, 
\author[e,f]{F.~Moia},
\author[p,b]{O.~Morra},
\author[n]{T.~Nagae}, 
\author[q]{H.~Outa},
\author[m]{A.~Pantaleo\thanksref{dead}},
\author[m]{V.~Paticchio}, 
\author[j]{S.~Piano}, 
\author[i,j]{R.~Rui},
\author[l,m]{G.~Simonetti},
\author[b]{R.~Wheadon}, \author[e,f]{A.~Zenoni}

\thanks[corresponding]{corresponding author. E-mail: filippi@to.infn.it; Phone:
+39.011.6707323; Fax:
+39.011.6707324.}
\thanks[dead]{Deceased}

\address[a]{Dipartimento di Fisica, Politecnico di Torino, Corso Duca degli
Abruzzi 24, Torino, Italy}
\address[b]{INFN Sezione di Torino, via P. Giuria 1, Torino, Italy}
\address[c]{Laboratori Nazionali di Frascati dell'INFN, via. E. Fermi, 40,
Frascati, Italy}
\address[d]{Department of Physics, Seoul National University, 151-742 Seoul,
South Korea}
\address[e]{Dipartimento di Ingegneria Meccanica e Industriale, 
Universit\`a di Brescia, via Branze 38, Brescia, Italy}
\address[f]{INFN Sezione di Pavia, via Bassi 6, Pavia, Italy}
\address[h]{Dipartimento di Fisica Sperimentale, Universit\`a di Torino,
via P. Giuria 1, Torino, Italy}
\address[ii]{SUBATECH, Ecole des Mines de Nantes, Universit\'e de Nantes,
CNRS-IN2P3, Nantes, France}
\address[i]{Dipartimento di Fisica, Universit\`a di Trieste, via Valerio 2,
Trieste, Italy}
\address[j]{INFN Sezione di Trieste, via Valerio 2, Trieste, Italy}
\address[k]{Dipartimento di Fisica Generale, Universit\`a di Torino,
via P. Giuria 1, Torino, Italy}
\address[ll]{CERN, CH-1211 Geneva 23, Switzerland}
\address[l]{Dipartimento di Fisica Universit\`a di Bari, via Amendola 173,
Bari, Italy}
\address[m]{INFN Sezione di Bari, via Amendola 173, Bari, Italy}
\address[n]{Department of Physics, Kyoto University, Kitashirakawa, Kyoto 
606-8502 Japan}
\address[o]{Department of Physics, Shahid Behesty University, 19834 Teheran,
Iran}
\address[p]{INAF-IFSI, Sezione di Torino, Corso Fiume 4, Torino, Italy}
\address[q]{RIKEN, Wako, Saitama 351-0198, Japan}

\begin{abstract}
The Non-Mesonic (NM) decay of $^4_\Lambda{\mathrm{He}}$
and $^5_\Lambda{\mathrm{He}}$  in two-body channels 
has been studied with the FINUDA apparatus. Two-body NM decays
of hypernuclei are rare and the existing observations and 
theoretical calculations are scarce and dated. 
The $^4_\Lambda{\mathrm{He}}\rightarrow d+d,\; p+t$ decay 
channels simultaneously observed by FINUDA  on several nuclei
are compared: the $pt$ channel is dominant. 
The decay yields for the two decay channels are assessed for the first time:
they are $(1.37\pm 0.37)\times 10^{-5}/K^-_{stop}$ and 
$(7.2\pm 2.7)\times 10^{-5}/K^-_{stop}$,
respectively. 
Due to the capability of FINUDA of identifying $^5_\Lambda{\mathrm{He}}$ 
hypernuclei, 
a few $^5_\Lambda{\mathrm{He}}\rightarrow d+t$ decay events
have also been observed. 
The branching
ratio for this decay channel has been measured for the first time: 
$(3.0\pm 2.3)\times 10^{-3}$.

\end{abstract}


\end{frontmatter}

\section{Introduction}
\label{intro}

Two weak decay modes of $\Lambda$ Hypernuclei are known, mesonic and  
Non-Mesonic (NM). The latter
gives a
unique tool to study the baryon-baryon weak interactions. 
Relatively
little information about NM decays 
can be inferred by alternative processes, such as nucleon
scattering experiments or parity violating nuclear transitions.

The NM
decay mode, in which two or more 
nucleons are emitted after  a $\Lambda$-$A$ multibaryon
weak interaction
in a nuclear medium, is dominant for medium-heavy nuclei. 
It  is characterized by a large momentum transfer; 
in fact, the emitted nucleons
have a  momentum in the range 400--600 MeV/$c$  
and can escape the nucleus.
In light hypernuclei
the NM decay
can also lead to final states composed by two bodies only: 
\begin{eqnarray}
^4_\Lambda\mathrm{He}&\rightarrow & d+d \\
^4_\Lambda\mathrm{He}&\rightarrow & p+t \\
^4_\Lambda\mathrm{He}&\rightarrow & n+\mathrm{^3He} \\
^5_\Lambda\mathrm{He}&\rightarrow & d+t. 
\end{eqnarray} 
Some of these reactions can be interpreted as two-nucleon induced decays.
FINUDA recently determined 
the two-nucleon induced NM decay
to be about 1/4 of the total NM, 
for $p$-shell hypernuclei
\cite{re:FND2NNM}. 
In addition, a 
suppressed branching ratio for the two-body NM channels 
is expected because of the large momentum transfer 
and of the possible two-step mechanisms involved.

Theoretical predictions are scarce due to the very limited available
data set. The only existing 
calculation for the two body $^4_\Lambda{\mathrm{He}}$ decay rates 
was performed in 1966 by
Rayet \cite{re:rayet}, and  was 
based on a phenomenological non-relativistic matrix
elements evaluation for the $\Lambda N\rightarrow N+N$ interaction. 
 Branching
ratios of 0.015 for (1) and (2) and of 0.03 for (3) were
predicted, with a spread within a 1.5 factor 
due to the dependence of the model
on the nuclear density and on the $\Lambda$ compression effect.

These evaluations are in rough agreement with the existing experimental
observations, which are scarce and dated. They all
belong to bubble chamber and emulsion experiments 
\cite{re:corenman,re:blockdd,re:keyes}. 
The full present database consists of a few 
$^4_\Lambda{\mathrm{He}}\rightarrow n + ^3\mathrm{He}$ events 
\cite{re:corenman,re:blockdd},
whose rate is 8--14\% of all the $^4_\Lambda{\mathrm{He}}$ NM decays,
and one $^4_\Lambda{\mathrm{He}}\rightarrow d+d$ event \cite{re:blockdd}. No
$^4_\Lambda{\mathrm{He}}\rightarrow p+t$ decays were observed.
Keyes {\it et al.} \cite{re:keyes}, in a liquid Helium bubble chamber 
experiment,
report 1.8\% as upper limit 
for the reactions (1) and (2), to be compared to the $\pi^-$ mesonic decay
of $^4_\Lambda{\mathrm{He}}$.  
For the $^5_\Lambda{\mathrm{He}}\rightarrow d+t$ decay only one event was
ever observed \cite{re:corenman}, and just one theoretical 
evaluation for the expected decay rates of 
$^5_\Lambda{\mathrm{He}}$  exists \cite{re:he5Theor}.

In this Letter the rates of the decay modes
(1) and (2) for  $^4_\Lambda\mathrm{He}$ hyperfragments 
and (4) for $^5_\Lambda\mathrm{He}$ 
are presented. The data were collected by the FINUDA spectrometer operating
at the DA$\Phi$NE 
$\phi$-factory,
Laboratori Nazionali di Frascati (LNF), Italy. 
FINUDA is a magnetic spectrometer designed for the study of 
hypernuclei production and decay 
being induced by stopped negative kaons 
on targets of different composition. 
The apparatus features a large geometrical acceptance 
($\sim 2\pi$ sr) and an
outstanding particle identification
capability for charged hadrons ($98\%$ and $94\%$ 
for protons and deuterons, respectively, integrated over the full
momentum range for their detection: 
(180--800) MeV/$c$ for protons, (300--800) MeV/$c$ for
deuterons). Also thanks 
to the apparatus transparency
an excellent momentum resolution is achieved,
up to  $\Delta p/p$  
$\sim 0.6\%$ for the
negative pions, of about 270 MeV/$c$, used in meson spectroscopy studies.
FINUDA cannot give any information about reaction (3), since 507 MeV/$c$
$^3$He nuclei cannot be detected by the apparatus.

The (1), (2) and (4) two-body NM decays of Hyperhelium isotopes
present a clear back-to-back topology 
for the reconstructed tracks. Such a feature, along with the PID information
provided by the apparatus, makes 
the full reconstruction of the hypernuclear decay products 
feasible.

The signature of two-body  $^4_\Lambda{\mathrm{He}}$  decays
is particularly clean. For reaction (1), it consists
of two 571.8 MeV/$c$ monochromatic back-to-back deuterons, for reaction
(2), the two hadrons $(p, t)$ are back-to-back emitted  
with a momentum of 508 MeV/$c$.

FINUDA can directly identify $^5_\Lambda{\mathrm{He}}$ hypernuclei and
indirectly detect $^4_\Lambda{\mathrm{He}}$ hyperfragments.
The identification of the 
$^5_\Lambda{\mathrm{He}}$ hypernucleus 
is performed by measuring negative pions from the
$K^- \;^6{\mathrm{Li}}\rightarrow {\mathrm{^6_\Lambda Li}}+\pi^-$ and
$K^- \; ^7{\mathrm{Li}}\rightarrow {\mathrm{^7_\Lambda Li}^\ast}+\pi^-$
reactions.
The unstable hypernuclei 
decay strongly to ${\mathrm{^5_\Lambda He}}$
according to the reactions
$^6_\Lambda{\mathrm{Li}}\rightarrow {\mathrm{^5_\Lambda He}}+ p$ and
$^7_\Lambda{\mathrm{Li}}^\ast\rightarrow {\mathrm{^5_\Lambda He}}+ d$. 
In these
reactions, the proton and the deuteron momentum 
is below 100 MeV/$c$, and their detection is not possible.
However, the formation pions 
have a momentum in a well defined range which 
allows for a clear $^5_\Lambda{\mathrm{He}}$ tagging.
Conversely, the 
$K^-_{stop}\, ^4{\mathrm{He}}\rightarrow {\mathrm{^4_\Lambda He}}+\pi^-$ 
reaction cannot be studied in FINUDA,
since a liquid Helium target could not be hosted in the experimental
set-up. 
 $^4_\Lambda{\mathrm{He}}$
hyperfragments can nevertheless be produced also
in the $K^-_{stop} A$ interaction in nuclei with
$A \geq 6$. 
In solid targets  the
$^4_\Lambda{\mathrm{He}}$ hyperfragment, however, 
cannot escape the target volume
and thus be 
tracked, as was done in emulsion experiments
\cite{re:lemonne,re:holland,re:schlein}. 
In these experiments, the measured yield of hyperfragment production in $K^-$
absorption at rest  
ranged from $(4.5\pm 0.5)\%$ \cite{re:sacton} to the
more recent values $(5\pm 1)\%$ \cite{re:davis} and $(6.5\pm 0.2)\%$
\cite{re:lemonne}. The fraction of the hyperfragment 
mesonic decay was measured to be
around 20\%. Hyperfragments were in general observed more copiously 
when produced by 
light emulsion nuclei, {\it i.e} up to $^{16}$O \cite{re:davis}.

\section{Outline of the FINUDA experimental apparatus}

A short description of the experimental set-up is given here for the sake of 
clarity. More details can be found, for instance, in Ref. \cite{re:FNDK2N}.
FINUDA features a cylindrical geometry and is installed in one of the
DA$\Phi$NE  interaction regions, where $\phi(1020)$ mesons from
the $e^+e^-$ collisions are produced. 
The charged kaons from 
$\phi(1020)\rightarrow K^++K^-$
decay (B.R.=0.49) have 
a momentum of 127 MeV/$c$ at
most, and they slow down while crossing the internal region of the apparatus
until they interact at rest in a set of targets.
The apparatus consists of five position-sensitive layers, arranged coaxially
around the beam axis. Four of them 
are also used for particle identification through
energy loss measurements. The tracking region is  
immersed in a uniform solenoidal magnetic field
of 1 T. Three main sectors 
may be singled out in the detector
layout: 
\begin{itemize}
\item interaction/target region: located at the apparatus center, consisting 
of a Beryllium beam pipe, 
a 12 scintillator slab hodoscope (TOFINO) \cite{re:tofino} used for
trigger purposes and for charged kaons discrimination, 
an eight module array of
double-side Si microstrip detectors (ISIM)
\cite{re:silici} 
facing  eight target tiles. 
The target set-up for the data used in the present analysis
consisted of two (90\% enriched) $^6$Li (thickness: 4 mm),
two natural isotopic composition
$^7$Li (4 mm), two $^9$Be (2 mm), one $^{13}$C (10 mm) 
and one $\mathrm{D_2O}$ (liquid filled and 
mylar walled, 3 mm thick) targets. 
\item tracking region: consisting of ten Si microstrip modules
(OSIM) \cite{re:silici} 
facing externally the targets, two arrays of eight
planar low-mass drift chambers (LMDC's) \cite{re:camere}, 
filled with a He-iC$_4$H$_{10}$ gas
mixture, and a  system of six 
longitudinal-stereo layers of
Ar-C$_2$H$_6$ filled straw 
tubes  \cite{re:straw}. 
The ISIM and OSIM modules feature a spatial resolution better than
$\sigma \sim 30\; \mu$m for both the $(r\phi)$ and $z$ coordinates. 
LMDC's provide a resolution $\sigma_{r\phi}\sim 150$ $\mu$m and 
$\sigma_z \sim 1$ cm, while for the straw tube system the resolution
is $\sigma_{r\phi}\sim 150$ $\mu$m and $\sigma_z \sim 500$ $\mu$m.
\item outer scintillator array: a barrel of 72 thick 
slabs \cite{re:tofone}, 
used for first level trigger, time-of-flight measurements with an
overall time resolution $\sigma\sim 800$ ps and
neutron identification with a $\sim 10\%$ efficiency.
\end{itemize}

The data used in the present analysis 
correspond to an integrated $e^+e^-$  luminosity of 
966 pb$^{-1}$,
collected by FINUDA in the 2006-2007 run.

\section{$^4_\Lambda{\mathrm{He}}\rightarrow d+d$ and 
$^4_\Lambda{\mathrm{He}}\rightarrow p+t$ decays}
\label{deut}

The $^4_\Lambda{\mathrm{He}}$ hyperfragment can be produced 
via the reaction
$K^-_{stop} \, ^A_ZX \rightarrow \; 
^4_\Lambda\mathrm{He}+\pi^-+ ^{A-4}_{Z-2}X'$,
where $X'$ is a recoiling 
nuclear system (bound or unbound), and the pion momentum is
larger than 220 MeV/$c$.
The hyperfragment production can also occur 
with the emission of a $\pi^0$, that however is
undetectable by FINUDA.

The signature of a $dd$ decay event consists of two high momentum 
back-to-back deuteron tracks.
The main source of background 
is given by the $^4_\Lambda{\mathrm{He}}\rightarrow (pn)+(pn)\pi^0$ 
mesonic decay, whose
frequency is almost comparable to the full NM branch \cite{re:He4Decay}. 
The mean momentum of the $(pn)$ pairs in this three-body decay is around
150 MeV/$c$; therefore, 
these events can be easily discarded by
applying a cut on the missing mass distribution of the
$^4_\Lambda{\mathrm{He}\rightarrow d+d}$  decay.

\begin{figure}[htb]
\begin{center}
\resizebox{\textwidth}{!}{\includegraphics{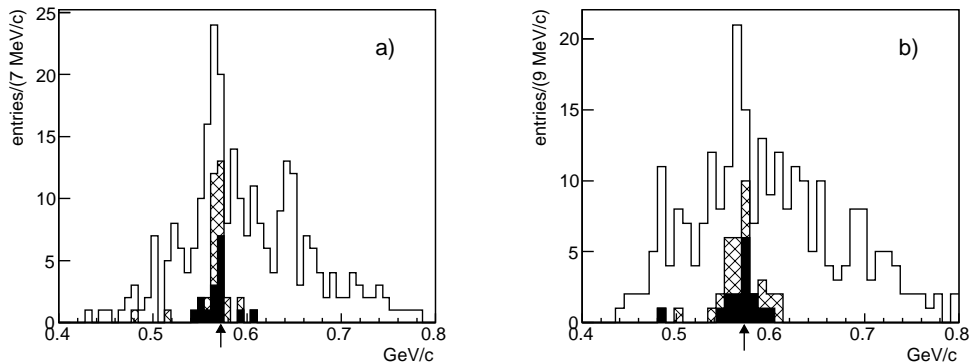}} 
\end{center}
\vspace{-0.5truecm}
\caption{Momentum distributions for the forward a) and the backward b)
deuterons in the two deuteron semi-inclusive sample. The hatched histograms
show $^4_\Lambda\mathrm{He}\rightarrow d+d$ at rest decay events. 
The black histograms correspond to $dd$ events with a high momentum
$\pi^-$ $(p_{\pi^-}>220$ MeV/$c$) in coincidence.} 
\label{fig:momdeut}
\end{figure}

The total inclusive $dd$ collected sample consists of 
$272\pm 16$ events, 
over all the FINUDA targets.
Fig. \ref{fig:momdeut} shows the momentum distribution of the two observed 
deuterons. In a pair, one of the deuterons is emitted 
from the target toward the outer hemisphere, {\it i.e.} toward
the tracking region: forward track, Fig. \ref{fig:momdeut}a). 
The second deuteron is 
emitted in the opposite direction, thus crossing twice ISIM, TOFINO, 
the beam pipe 
and one of the targets: backward track, Fig. \ref{fig:momdeut}b). The forward
tracks are required to be reconstructed by a minimum of three hits, one 
of which is necessarily located on the OSIM array. 
The momentum
resolution of backward deuterons is spoilt by the 
larger material budget crossed
by the particle, but 
at least 
two hits of the track are located on the high resolution ISIM and OSIM
detectors. 
From Monte Carlo simulations 
the momentum resolution 
of forward deuterons
in FINUDA results to be 3\% FWHM (17 MeV/$c$ at 570 MeV/$c$), 
while for backward deuterons is 4\% FWHM ($\sim 22$ 
MeV/$c$). Both tracks were selected with a 
track fitting procedure 
which requires $\chi^2<20$ (C.L. $\sim 95\%$). 

\begin{figure}[htb]
\begin{center}
\resizebox{\textwidth}{!}{\includegraphics{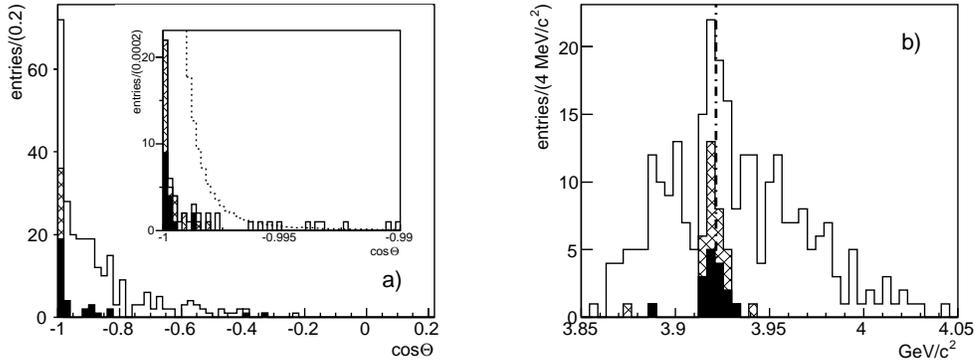}} 
\end{center}
\caption{Distribution of the angle between the two deuterons (a) and 
of the $(dd)$ invariant mass (b). Open histogram: semi-inclusive
$dd$ sample. The hatched histograms
display $^4_\Lambda\mathrm{He}\rightarrow d+d$ at rest decay events. 
The black histograms correspond to $dd$ events with a high momentum
$\pi^-$ $(p_{\pi^-}>220$ MeV/$c$) in coincidence.
The inset in Fig. a) zooms the distribution for $\cos\theta<-0.99$ angles.
The dotted histogram in the inset, normalized to the tail of the
experimental histogram, shows the corresponding distribution
for simulated $^4_\Lambda{\mathrm{He}}\rightarrow d+d$ at rest decays,
filtered through the detector acceptance.
In Fig. b) the dot-dashed line marks the
$^4_\Lambda{\mathrm{He}}$ mass value.
}
\label{fig:ang_im}
\end{figure}

The semi-inclusive momentum
distributions are shown as open histograms in Fig. \ref{fig:momdeut}. 
Fig. \ref{fig:ang_im}a) shows
the distribution of the angle between the two deuteron tracks, 
while Fig. \ref{fig:ang_im}b)
displays  the invariant mass distribution 
of the $dd$ pairs. 
Events 
with a sharp back-to-back angular correlation $(\cos\Theta<-0.995$)
are chosen, see inset of Fig. \ref{fig:ang_im}a). 
The dotted histogram in the inset shows the corresponding distribution
for simulated $^4_\Lambda{\mathrm{He}}\rightarrow d+d$ at rest decays,
filtered through the detector acceptance and normalized to the tail of
the experimental distribution.
The applied cut allows all the searched events to be accepted 
eliminating  $dd$ non-monochromatic pairs emitted in possible
heavier hyperfragment production and decay.

In the back-to-back $dd$ sample, exclusive 
events are then required to have a total energy in 
the (3--4) GeV range and 
a total momentum 
less than 50 MeV/$c$. 
A last requirement imposes the
missing mass for the $^4_\Lambda{\mathrm{He}}\rightarrow d+d$ decay to 
be compatible with zero within
a $2\sigma$ range ($\sigma = 11$ MeV/$c^2$).
The events selected for the final yield evaluation are shown in 
the cross-hatched  
distributions in Fig. \ref{fig:momdeut} and \ref{fig:ang_im}.

Summing over all the targets,
the total statistics available after the described selections 
is $31\pm 6$ $dd$ events. 
$14\pm 4$ events, out of the semi-inclusive $dd$ sample, 
present an additional $\pi^-$ with
momentum larger than 220 MeV/$c$. These events are displayed in 
the black histograms
in Fig. \ref{fig:momdeut} and \ref{fig:ang_im}.
They
can be interpreted as exclusive events in which both the  
$^4_\Lambda{\mathrm{He}}$ hyperfragment formation and its decay have been
measured. However, they are not used for yield evaluations as not all of
them present strictly back-to-back deuterons, as shown in Fig. 
\ref{fig:ang_im} a).

The selected sample is affected by a background contribution 
consistent with zero. Side bin background 
evaluations have been made selecting different
missing mass and momentum ranges, displaced one or two $\sigma$'s
from the respective reference values.
No events matching the required event signature 
are found in displaced ranges. 

A target-by-target measurement of the number of 
$^4_\Lambda{\mathrm{He}}\rightarrow d+d$ decays at rest per incident 
$K^-_{stop}$ can be performed. The recorded
number of collected $K^-_{stop}$ 
per target depends on the target position, due to the slight
$\phi$ boost, and it varies in the
range (1.1--2.1)$\times 10^7$.
The results are reported, for different nuclei, 
in Tab. \ref{tab:ddY}. The yield is basically constant for
lighter targets (up to $^9\mathrm{Be}$); for heavier targets
 an upper limit can be given 
at 90\% confidence level. The overall efficiency, which takes into
account the trigger, the apparatus acceptance, the reconstruction
and analysis procedures as well as the efficiency of the
detectors, is also a function of the target position and varies in the 
range (1.5--3)$\%$.  
The systematic errors take into account the spread of the 
values measured in
different targets of the same composition and the effect of
1$\sigma$ variations in the selection criteria. 
They also take into account the 
uncertainties in assessing the number of stopped kaons 
(due to out of target and in-flight 
interactions, $K^-/K^+$ swap in the pattern recognition procedure, 
backtracking algorithm inefficiencies).

\begin{table}[h]
\centering
\resizebox{\textwidth}{!}{
\begin{tabular}{||c||c|c||c|c||}
\hline
target & $dd$ Events & Yield $\times 10^{-5}/(K^-_{stop})$ & 
$pt$ Events & Yield $\times 10^{-5}/(K^-_{stop})$ \\
\hline
\multirow{2}{*}{$^6\mathrm{Li}$} & $12\pm 3$ & $3.0\pm 1.3_{stat}\pm 0.9_{sys}$ &
$1\pm 1$  & $5.0\pm 6.5_{stat}\pm 0.3_{sys}$   \\
\ & \ & \ & \ &  $<16.8 \;\; (90\%{\mathrm{C.L.}})$ \\ \hline
\multirow{2}{*}{$^7\mathrm{Li}$} & $7\pm 3$ & $2.4\pm 1.3_{stat}\pm 0.8_{sys}$ 
& $1\pm 1$ &  $5.1\pm 5.8_{stat}\pm 0.3_{sys}$ \\
\ & \ & \ & \ & $<14.3 \;\; (90\%{\mathrm{C.L.}})$\\ \hline
$^9\mathrm{Be}$ & $10\pm 3$ & $3.3\pm 1.4_{stat}\pm 0.4_{sys}$ 
& $5\pm 2$ & $14.9\pm 7.4_{stat}\pm 0.9_{sys}$\\ \hline
\multirow{2}{*}{$^{13}\mathrm{C}$} & $1\pm 1$ & $0.6\pm 0.6_{stat} \pm 1.0_{sys}$ &
$1\pm 1$ & $7.2\pm 7.8_{stat}\pm 0.2_{sys}$ \\ 
\ & \ & $< 2.3\;\; (90\%{\mathrm{C.L.}})$ & \ & $<30.5 \;\; (90\%{\mathrm{C.L.}})$\\ \hline
\multirow{2}{*}{$^{16}\mathrm{O}$} & $1\pm 1$ & $0.6\pm 0.6_{stat}\pm 1.1_{sys}$
& $2\pm 1$ & $10.4\pm 8.0_{stat}\pm 0.2_{sys}$\\
\ & \ & $< 2.7 \;\; (90\%{\mathrm{C.L.}})$ & \ & \ \\
\hline
\end{tabular}
}
\caption{
Second and third column, number of events and
yields, per $K^-_{stop}$,  of 
the $^4_\Lambda{\mathrm{He}}\rightarrow d+d$ decay, for hyperfragments produced
all targets of given $A$ (first column).
Fourth and fifth column, number of events and
yields per $K^-_{stop}$ of the 
$^4_\Lambda{\mathrm{He}}\rightarrow p+t$ decay at rest.
In case of one event only, an upper limit following Poisson statistics is
also quoted.}
\label{tab:ddY}
\end{table}

An overall average yield can be deduced.
The resulting value is 
$Y_{^4_\Lambda\mathrm{He}\rightarrow d+d}=(1.37 \pm 0.37)
\times 10^{-5}/K^-_{stop}$, where the overall 
uncertainty accounts for
statistical and systematic errors added in quadrature.
Assuming a hyperfragment production rate of about $5\%/K^-_{stop}$, 
 roughly valid up to $A=16$ \cite{re:davis},
and under the simple hypothesis that 
$^4_\Lambda{\mathrm{He}}$'s are the most abundantly produced hyperfragments,
this yield corresponds to an upper limit for the
 $^4_\Lambda{\mathrm{He}}\rightarrow d+d$ branching ratio of
about $3\times 10^{-4}$.

Concerning the $^4_\Lambda{\mathrm{He}}\rightarrow p+t$ decay channel,
tritons in FINUDA can be detected
only for momenta larger than 550 MeV/$c$, which exceeds 
the momentum value expected for the
two body $^4_\Lambda{\mathrm{He}}\rightarrow p+t$ decay at rest,
508 MeV/$c$.
Tritons with lower momenta, in fact, cannot
be reconstructed as they stop in the inner tracking layers of FINUDA. 
However, if  they cross the I/OSIM layers they deposit
a large amount of energy, 
that can be
used to tag such events. 
As a consequence, the $^4_\Lambda{\mathrm{He}}\rightarrow p+t$ decay
 can effectively be measured only 
through the information on the proton 
and the momentum of the formation pion. The proton track is 
selected applying the same quality criteria described above
for deuterons.
For the pion the quality criteria are released.
A cut is applied 
to eliminate
contributions from Quasi-Free (QF) 
reactions due to the absorption of the $K^-$ by just 
one nucleon: $p_{\pi^-} > 200$ MeV/$c$. On the other hand,
pions with
$p_{\pi^-} > 220$ MeV/$c$
are expected from the observation of the coincident $\pi^-$'s
in the analysis of the $^4_\Lambda{\mathrm{He}}\rightarrow d+d$
decay. 
Events with
a detected neutron are moreover rejected to
remove possible  
contaminations due to baryon ($\Lambda,\,\Sigma^-$) decays.
A total sample of $297023 \pm 545$ semi-inclusive $(p\pi^-)$ events is 
collected.
None of the events with a proton selected in the proper 
momentum band (498--520 MeV/$c$)
feature hits  with
large energy deposit in silicon layers opposite to the emitted proton track.

\begin{figure}[htb]
\begin{center}
\resizebox{9truecm}{!}{\includegraphics{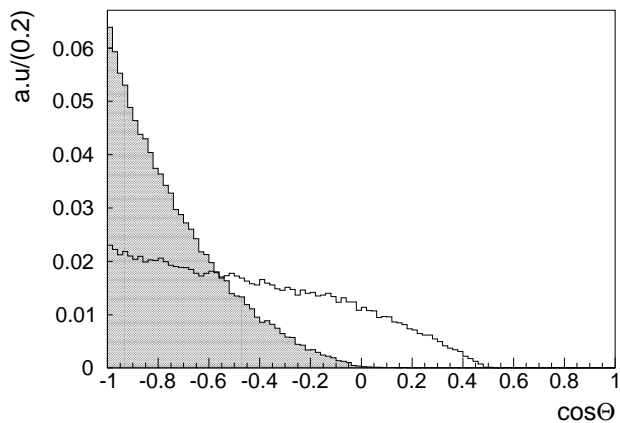}} 
\end{center}
\caption{
Monte Carlo data: distribution of the angle formed between a 
proton and a $\pi^-$ track in the simulation of the $QF$
$K^-(pn) \rightarrow \Sigma^- p$ (grey histogram), and of the 
formation of a $^4_\Lambda{\mathrm{He}}$ hyperfragment and 
its decay at rest in the $p+t$ channel (open histogram).
The two distributions are arbitrarily normalized; the simulated
data are modulated by the apparatus acceptance.
}
\label{fig:angolo}
\end{figure}

The main background source for the $^4_\Lambda{\mathrm{He}}\rightarrow p+t$ 
decay channel is played by the QF 
$K^-(pn)\rightarrow\Sigma^- +p$ two nucleon absorption ($K2N$), 
which has a   
large capture rate, $1.62\%/K^-_{stop}$ in $^6$Li \cite{re:FNDK2N}
and an order of magnitude larger in heavier nuclei. The QF
signature is similar to that of the studied decay,  
with a ``quasi''-mo\-no\-chro\-ma\-tic proton of about 510
MeV/$c$ momentum, and a fast $\pi^-$. 
A study of the impact parameters and the angular
distributions of the emitted proton and negative pion has been performed
to understand how to
remove the most sizeable fraction of $K2N$-QF events without 
loosing too much of the searched signal.  
An optimized cut in the
distribution of the
$\pi^-$ impact parameter, {\it i.e.} the distance between the track and
the $K^-$ interaction vertex, 
rejects about 79\% of
the $\Sigma^- p$ background, but it also reduces by 60\% the 
searched signal. 
The $(p\pi^-)$ angle in the
$K^-(pn)\rightarrow \Sigma^- +p$ absorption displays a marked 
back-to-back trend. The simulated distributions of the $(p\pi^-)$ angle
for particles emitted in the $K2N$ reaction (grey area) or from 
$^4_\Lambda\mathrm{He}$ formation and decay (open histogram)
are shown in Fig. 
\ref{fig:angolo}. 

The distributions are arbitrarily normalized and are filtered through 
the apparatus acceptance, which suppresses the positive cosine region for 
both the reactions. Still, it emerges  that, due to the apparatus acceptance,
in the case of hypernuclear formation
and decay
one should expect a relatively enhanced
emission of tracks on the same side with respect to the target. 
The selection of such tracks
also enhances the contamination of 
$\Lambda$'s, though, 
whose contribution can be removed by a proper cut on the  
$p\pi^-$ invariant mass.
The selection of $p\pi^-$ pairs emitted in the same
hemisphere reduces the
available sample of 53\%, but the $\Sigma^-p$ contamination is 
reduced to 12\%. This selection criterion is therefore chosen.

\begin{figure}[htb]
\begin{center}
\resizebox{\textwidth}{!}{\includegraphics{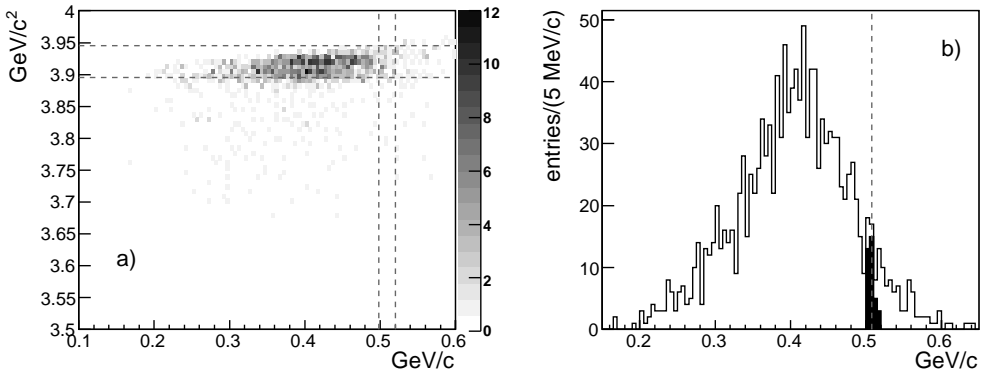}} 
\end{center}
\caption{a) Proton momentum vs missing mass of the reaction
$K^- \, ^A_ZX \rightarrow \; \pi^- + Hyp  + ^{A-4}_{Z-2}X'$. The horizontal
band between the two dashed lines corresponds to the interval
$m(^4_\Lambda\mathrm{He})\pm 25$ MeV/$c^2$; the vertical band corresponds
to the momentum interval $508\pm 12$ MeV/$c$ across the nominal
proton momentum value
for the $^4_\Lambda\mathrm{He}\rightarrow p+t$ decay at rest.
b) Proton momentum distribution for all events with a high momentum
coincident pion (open histogram), and for events selected in the horizontal
band of a) having a missing mass for the 
$^4_\Lambda\mathrm{He}\rightarrow p+t_{miss}$ decay
in the range $(m_t\pm 5)$ MeV/$c^2$. The vertical dashed line
marks the nominal proton momentum 
for the $^4_\Lambda\mathrm{He}\rightarrow p+t$ 
decay at rest.
}
\label{fig:ptspectra}
\end{figure}

The events surviving the mentioned selections fill the 
histograms in 
Fig. \ref{fig:ptspectra}a) and b).
The scatter plot in Fig. \ref{fig:ptspectra}a) shows the
proton momentum versus the missing mass of the reaction
$K^- \, ^A_ZX \rightarrow \; \pi^- + Hyp + ^{A-4}_{Z-2}X'$,
where the $Hyp$ system is, presumably, a hyperfragment.
In this reaction the $X'$ and  $Hyp$ systems are not measured,
and $X'$ is assumed to be a spectator. The mass of $X'$ is
assumed to be equal to the mass
of the most stable nuclear system with $(A-4)$ mass number and 
$(Z-2)$ protons.

The horizontal band between the dashed lines
is centered at the $^4_\Lambda{\mathrm{He}}$ 
mass $(m=3.92$ GeV/$c^2$) and is  50 MeV/$c^2$ wide. 
The vertical band is, conversely,
centered at the nominal proton momentum for the 
$^4_\Lambda{\mathrm{He}}\rightarrow p+t$ decay, and is 24 MeV/$c$
wide. The highest intensity of the plot is located around $p_p = 420$ MeV/$c$
(one-nucleon induced NM weak decay).
At the intersection of the two bands a small activity can be noticed.
The open histogram in Fig. \ref{fig:ptspectra}b) shows the
x-axis projection of this scatter plot.

The events in the horizontal band of the scatter plot can be
selected and for them the missing mass distribution of the
$Hyp\rightarrow p+t_{mass}$ reaction can be studied (figure not shown).
This distribution has a spread of some tens of MeV due to the
hypotheses applied for the kinematics and the mass
of the unmeasured recoiling
nuclear system. For this reason, a selection on the measured
deuteron momentum is preferred requiring,
target by target, the missing mass of the 
$^4_\Lambda\mathrm{He}\rightarrow p+t_{miss}$ at rest decay to 
be in the range $(m_t\pm 5)$ MeV/$c^2$: the selected
events are shown in the black histogram in Fig. \ref{fig:ptspectra}b),
which is centred at 508 MeV/$c$ with a width
of 15 MeV/$c$. The momentum resolution for such a proton  
is 0.7\% ($\sigma$).
The events in the black histogram are then used for the final evaluation
of the yields.
 
\begin{figure}[htb]
\begin{center}
\resizebox{\textwidth}{!}{\includegraphics{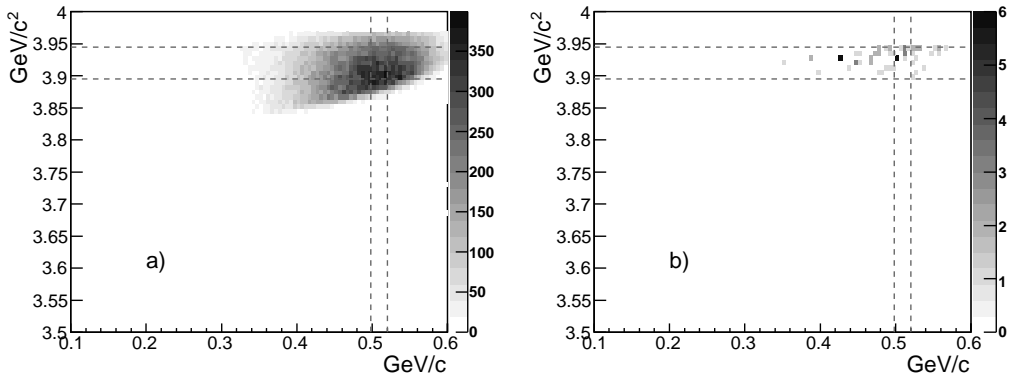}}
\end{center}
\caption{Scatter plots of the proton momentum vs the 
missing mass of the reaction
$K^- \, ^A_ZX \rightarrow \; \pi^- + XX  + ^{A-4}_{Z-2}X'$ for
events simulated in the $K^-(pn)\rightarrow \Sigma^-p$ 
background channel: a) no selection cuts applied;
b) selection cuts applied as for real data.
}
\label{fig:sigmaPBck}
\end{figure}

The selected sample, still background inclusive,
 amounts to a total of $44\pm 7$ $pt_{miss}\pi^-$
events (from all available targets). The contamination in
the selected sample due to the $\Sigma^-p$ source consists of a few
$10^{-5} \mathrm{\; events}/K^-_{stop}$, being so, presumably, of 
the same order of magnitude of 
the yield of the searched signal. In the final sample 
the signal to background ratio
amounts to $S/N=0.23$, which corresponds to a 2.9$\sigma$ 
statistical significance of the observed signal.
The scatter plot of Fig. \ref{fig:sigmaPBck}a) shows the same distribution
presented in Fig. \ref{fig:ptspectra}a) for 
$K^- (pn)\rightarrow \Sigma^- p$ simulated 
events, without any kinematic cut. The
distribution covers completely the region of the searched signal,
indeed it  has its maximum close to the
intersection of the chosen bands. However, if
the described angular selection is applied to these data,
the distribution in  Fig. \ref{fig:sigmaPBck}b) is obtained,
which shows that a significant reduction of the background contribution
can be achieved, even if an unavoidable contamination is 
still present in the region under study.

After background subtraction, performed target-by-target by side-bin 
evaluations in the proton momentum spectra,
an assessment of the yield of the 
$^4_\Lambda{\mathrm{He}}\rightarrow p+t$ decay  is done and the results
are reported in columns 5 and 6 of 
Tab. \ref{tab:ddY}. An overall efficiency for this channel, including 
acceptance, trigger,
reconstruction, analysis cuts as well as detector failures, is estimated to
be around $1.4\times 10^{-3}$ (modulated according to the target position).
In case of a single selected event 
 an upper limit, at 90\% C.L., may be given. 
The quoted systematic uncertainties
take into account the maximum spread of values obtained with small
variation of the selection cuts and in different
targets of the same nuclear composition.
They also take into account the uncertainty 
on the kaon normalization described in the previous section.

From an examination of the obtained yields, the
$^4_\Lambda{\mathrm{He}}\rightarrow p+t$ decay at rest looks favoured
in  heavier targets. 
Following the method described above for $dd$ decays, the average value
$(7.2\pm 2.7)\times 10^{-5}/K^-_{stop}$ is obtained.
Assuming again a hyperfragment production 
of about $5\%/K^-_{stop}$, the upper limit for the
branching ratio of the $^4_\Lambda{\mathrm{He}}\rightarrow p+t$ decay is
  $\sim 1.4\times 10^{-3}$.

\begin{figure}[htb]
\begin{center}
\resizebox{9truecm}{!}{\includegraphics{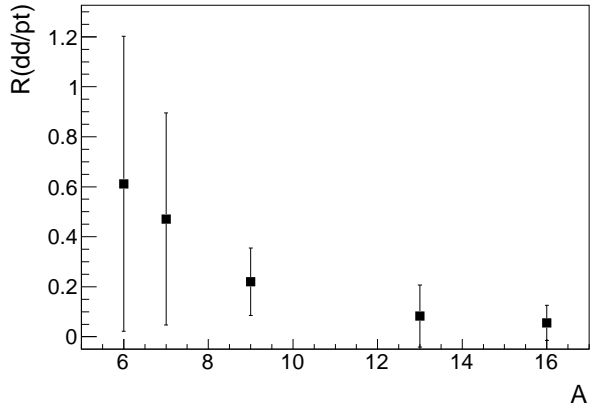}} 
\end{center}
\caption{Ratio of the $^4_\Lambda{\mathrm{He}}\rightarrow d+d$
to $^4_\Lambda{\mathrm{He}}\rightarrow p+t$ decay yields 
as a function of the
atomic mass number $A$.}
\label{fig:ratio}
\end{figure}

The ratio between the absolute number of events 
collected in the $dd$ and $pt$ 
channels can also be evaluated target by
target.
The trend of the ratio 
as a function of $A$ is reported in Fig. \ref{fig:ratio}, 
in which
the error bars take into account both statistic and systematic errors. 

The weighted average value over the (6--16) $A$ range is  
$R(^4_\Lambda{\mathrm{He}}\rightarrow d+d/
^4_\Lambda{\mathrm{He}}\rightarrow p+t) = (9.9\pm 5.5)\times 10^{-2}$. 
In spite of the 
large errors, a dominance of the $pt$ decay mode over the $dd$ 
arises, in particular for nuclei with small $A$.

\section{$^5_\Lambda{\mathrm{He}}$ production and 
$^5_\Lambda{\mathrm{He}}\rightarrow d+t$ 
decay in FINUDA}

FINUDA can observe both $^5_\Lambda{\mathrm{He}}$ hypernuclei and
hyperfragments, produced in targets heavier than Li.
In FINUDA $^5_\Lambda{\mathrm{He}}$ hypernuclei may be produced via the
$K^-_{stop}$$^{6,7}$Li
interaction. 
They can be observed by FINUDA
in the formation $\pi^-$ momentum spectrum by requiring a proton from 
the non-mesonic decay of the
hypernucleus in coincidence \cite{re:FNDNM}. 

The data set for the $^5_\Lambda{\mathrm{He}}\rightarrow dt$ decay is 
obtained  
by selecting a negative pion in the momentum
range 267--273 MeV/$c$ along with a  
deuteron in the 570--630 MeV/$c$ interval;
the deuteron from the decay
has a nominal momentum of 597 MeV/$c$. 
Still, this selection 
prevents further tritons from being detected. But
events with a large energy release on ISIM or OSIM 
modules opposite to the emitted deuteron track have been found: 
one event with all the above features 
has been reconstructed for
$^6$Li 
and  two for $^7$Li. 
No events presenting
the same topology and momenta slightly outside the
mentioned momentum ranges have been found, so the background contribution
to these events
is consistent with zero.
With an overall efficiency of about (6--11)$\times 10^{-3}$,
the yield  
$Y_{^5_\Lambda{\mathrm He}\rightarrow d+t}=
(2.6\pm 1.5_{st}\pm 1.1_{sys})\times 10^{-5}/K^-_{stop}$ is found, 
with
a statistical significance of 1.7$\sigma$. 
Normalizing to 
the absolute number of $^5_\Lambda\mathrm{He}$ hypernuclei measured by
FINUDA in the same run \cite{re:FNDNM}, 
the branching ratio for
this decay channel can be evaluated:
$B.R.({^5_\Lambda\mathrm{He}\rightarrow d+t}) = 
(3.0 \pm 2.3)\times 10^{-3}$.
On the same data sample, an assessment of the one-proton induced NM decay 
rate (inclusive of FSI distortion effects) has been determined,
$R({^5_\Lambda\mathrm{He}}) = 0.28 \pm 0.07$
\cite{re:FNDNM}. 
Taking also into account the non-measured one-neutron
and two-nucleon induced decays,  
the evaluated $dt$ branching ratio is about two orders of magnitude smaller
than the NM decay rate,
roughly in agreement 
with the theoretical expectations \cite{re:rayet}. 
The same conclusion can also be attained by deducing the NM branching ratio
from the total and mesonic widths of the ${^5_\Lambda\mathrm{He}}$ decays
\cite{re:kameoka}.

A second analysis of the $^5_\Lambda{\mathrm{He}}\rightarrow dt$ decay is 
based on the search of $^5_\Lambda{\mathrm{He}}$ hyperfragments, following
the same analysis pattern applied in  the $^4_\Lambda{\mathrm{He}}$ case.
A total sample of 
$1056 \pm 32$ $d\pi^-$ inclusive events is available. 
Such events rely on the measurement of a 
  high-resolution fast $\pi^-$ 
($\Delta p/p \sim 0.75\%,\; p_{\pi^-}>200$ MeV/$c$) in coincidence with a 
deuteron. 
A stringent $\chi^2$ from of the
track fitting procedure is required, corresponding to C.L. $\sim 98\%$. 
The resolution for a deuteron of $\sim$ 600 MeV/$c$ momentum is 8 MeV/$c$
$(\sigma)$, corresponding to $\Delta p/p \sim 1.5\%$.

The  $K^- \, ^A_ZX \rightarrow \; \pi^- + Hyp + ^{A-5}_{Z-2}X'$
events, whose missing mass falls in a region centered at the
$^5_\Lambda\mathrm{He}$ mass (4.85 GeV/$c^2$), 40 MeV/$c^2$ wide,
are further studied. In this reaction the recoiling nuclear
system is supposed to be produced at rest as a spectator.
Furthermore, 
the $^5_\Lambda\mathrm{He}$ hyperfragment is assumed to be produced by proton 
emission from $^6_\Lambda\mathrm{Li}$ if the $K^-$ is stopped in $^6$Li,
while it is assumed to be produced by deuteron emission from 
$^7_\Lambda\mathrm{Li}^\ast$ when $K^-$'s are stopped in other targets.
This assumptions change slightly the reaction kinematics for different targets.

\begin{figure}[htb]
\begin{center}
\resizebox{\textwidth}{!}{\includegraphics{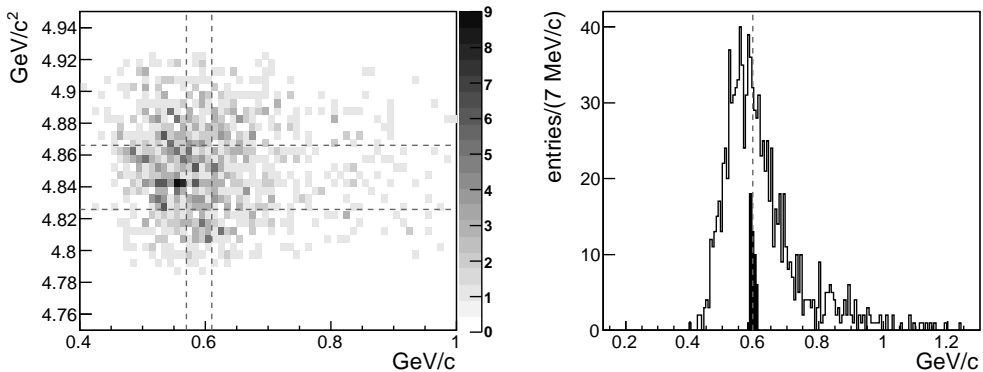}} 
\end{center}
\caption{a) Scatter plot of the deuteron momentum vs the missing mass
of the  $K^- \, ^A_ZX \rightarrow \; \pi^- + Hyp + ^{A-5}_{Z-2}X'$
reaction. The horizontal
band between the two dashed line corresponds to the interval
$m(^5_\Lambda\mathrm{He})\pm 20$ MeV/$c^2$; the vertical band corresponds
to the momentum interval $597\pm 20$ MeV/$c$ across the nominal
deuteron momentum value
for the $^5_\Lambda\mathrm{He}\rightarrow d+t$ decay at rest.
b) Deuteron momentum distribution for all events with a high momentum
coincident pion (open histogram), and for events selected in the horizontal
band of a) so that the missing mass
$^5_\Lambda\mathrm{He}\rightarrow d+t_{miss}$ 
at rest decay is
in the range $(m_t\pm 12.5)$ MeV/$c^2$. The vertical dashed line
marks the nominal deuteron momentum for the 
$^4_\Lambda\mathrm{He}\rightarrow p+t$ 
decay at rest.
}
\label{fig:dtspectra}
\end{figure}

The scatter plot of the deuteron momentum versus the 
missing mass of the above reaction
is displayed in Fig. \ref{fig:dtspectra}a). The horizontal
band between the two dashed lines shows the range selected 
around the $^5_\Lambda\mathrm{He}$ mass; the spread within the band
is again due to the hypothesis made on the kinematics of the recoiling
nuclear system. The vertical
band centered at 597 MeV/$c$, 40 MeV/$c$ wide, marks the deuteron momentum
range of interest. In the intersection region some limited activity  
can be seen.
The events in the horizontal band may be further selected asking the 
missing mass of the $^5_\Lambda\mathrm{He}\rightarrow d+t$ at rest decay 
to belong, target by target, 
to the interval $(m_t\pm 12.5)$ MeV/$c^2$ (equivalent to a cut on the
deuteron momentum). 
Events surviving this selection are shown in the black histogram 
in Fig. \ref{fig:dtspectra}b), where the deuteron momentum is displayed
for inclusive events also (open histogram). These events are 
located in a range, 
30 MeV/$c$ wide, across 597 MeV/$c$.

\begin{figure}[htb]
\begin{center}
\resizebox{9truecm}{!}{\includegraphics{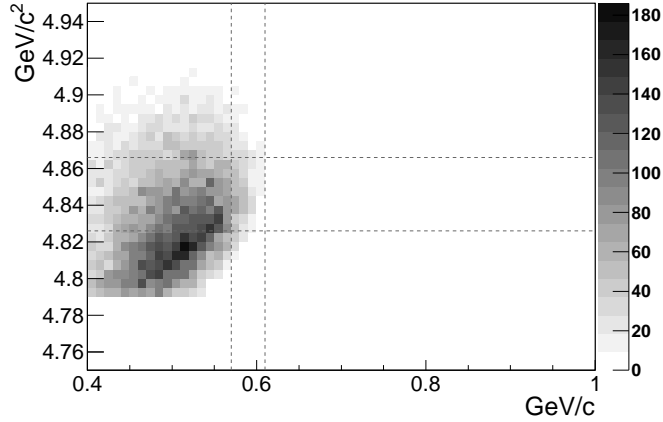}}
\end{center}
\caption{Scatter plot of the proton momentum vs the 
missing mass of the reaction
$K^- \, ^A_ZX \rightarrow \; \pi^- + XX  + ^{A-5}_{Z-2}X'$ for
events simulated in the $K^-(pn)\rightarrow \Sigma^-p$ 
background channel, assuming that the proton is misidentified as a 
deuteron; the events are then subject to the same  analysis chain 
used to
identify the $^5_\Lambda{\mathrm{He}}\rightarrow dt_{miss}$ decay.
}
\label{fig:sigmaP_DTBck}
\end{figure}

Also the background contribution in 
the $\pi^-dt_{miss}$
sample is sizeable and mostly due, again, to the 
$K^-(pn)\rightarrow \Sigma^- +p$ QF reaction, where the $p$ is
misidentified as $d$ (which can occur at the level of a few
percent). 
Fig. \ref{fig:sigmaP_DTBck} shows the scatter plot of the momentum vs the
missing mass for inclusive $d\pi^-$ events from a simulated
$K^-(pn)\rightarrow \Sigma^-p$ sample, selected forcing
a proton/deuteron misidentification.
A non-negligible background contribution still remains in the
signal region. It has been subtracted from the signal, target by target,
resorting to side-bins counting in the deuteron momentum 
distribution.
The $S/N$ ratio, averaged over all the available targets, is  
about 20\%.
A further source of background could arise from the $K^-$ absorption by
three nucleons with deuteron emission. However, this contribution is
neglected in this analysis as the reaction is expected to
occur at a rate at least one order of magnitude smaller than two-nucleon
absorption \cite{re:lambdaD}.

Tab. \ref{tab:dtY} reports
the decay rate per $K^-_{stop}$  of 
$^5_\Lambda\mathrm{He}$ hyperfragments
in the $dt$ channel. The systematic uncertainty takes 
into account the spread of the values due to 1$\sigma$ changes 
in the selection 
criteria, the different counting in targets of the same nucleus type, 
as well as the correction factors in the kaon normalization.
The overall efficiency is in the range (3--6)$\times 10^{-3}$.
The yield averaged over the available nuclei is 
$(1.40\pm 0.24)\times 10^{-4}/K^-_{stop}$.

\begin{table}[h]
\centering
\begin{tabular}{||c|c|c||}
\hline
target & Number of events & Yield$\times 10^{-4}/(K^-_{stop})$ \\
\hline
$^6\mathrm{Li}$ & $ 4\pm 2$ & $1.83\pm 0.93_{stat}\pm 0.12_{sys}$ \\
$^7\mathrm{Li}$ & $ 5\pm 2$ & $1.12\pm 0.51_{stat}\pm 0.08_{sys}$ \\
$^9\mathrm{Be}$ & $ 13 \pm 4$ & $1.23\pm 0.38_{stat}\pm 0.02_{sys}$ \\
$^{13}\mathrm{C}$ & $ 7 \pm 3$ & $2.25 \pm 0.87_{stat}\pm 0.04_{sys}$ \\
$^{16}\mathrm{O}$ & $ 11 \pm 3$ & $1.58\pm 0.50_{stat}\pm 0.03_{sys}$ \\
\hline
\end{tabular}
\caption{Yield of the $^5_\Lambda{\mathrm{He}}\rightarrow d+t_{miss}$ decay at
rest per $K^-_{stop}$.}
\label{tab:dtY}
\end{table}
 
\section{Summary}
FINUDA has been able to measure several features of the two-body 
decay channels $^4_\Lambda{\mathrm{He}}\rightarrow d+d$,
$^4_\Lambda{\mathrm{He}}\rightarrow p+t$ and
$^5_\Lambda{\mathrm{He}}\rightarrow d+t$,
thus conveying
new results useful to 
complete the meager existing database on two-body non-mesonic decay
modes of light hypernuclei.
Despite  the limited statistics, FINUDA has observed
signatures of decay events with a reduced background
in some of the studied channels. The particle identification
capabilities of the magnetic spectrometer
together with its high resolution allow for 
the detection of a few rare events
never observed before.

For $^4_\Lambda{\mathrm{He}}\rightarrow d+d$
in different targets
the decay yields 
are of the order of a few
$10^{-5}/K^-_{stop}$, while for  the $pt$ decay channel 
they are some units larger. 
The $^5_\Lambda{\mathrm{He}}\rightarrow d+t$
decay yield of $^5_\Lambda{\mathrm{He}}$ 
hypernuclei is
$(2.6\pm 1.5_{st}\pm 1.1_{sys})\times 10^{-5}/K^-_{stop}$.  
The B.R. for this decay channel
has been determined 
for the first time to be $(3.0\pm 2.3)\times 10^{-3}$. 
This is in rough agreement 
with the theoretical prediction \cite{re:rayet} of 
two orders of magnitude less as compared to 
the one-nucleon induced NM decay branching ratio \cite{re:FNDNM}.

\end{document}